\documentclass[showpacs,showkeys,a4paper,nofootinbib,superscriptaddress]{revtex4-2}
\usepackage{graphicx}
\graphicspath{ {./images/} }

\usepackage{newpxmath,newpxtext}
\usepackage{graphicx}
\usepackage{epsfig}
\usepackage[normalem]{ulem}
\usepackage{xcolor}

\newcommand{\pder}[2]{\dfrac{\partial#1}{\partial#2}}

\newcommand{\pdot}[1]{\dot{\partial}_{#1}}

\newcommand{\Gd}{\mathcal{G}}

\newcommand{\R}{\mathcal{R}}
\newcommand{\de}{\mathrm{d}}

\newcommand{\lin}{\\[7pt]}
\usepackage{tensor}

\begin{document}

\title{Cosmology Based on  Finsler and Finsler-like Metric Structure of Gravitational Field}

\author{P. C. Stavrinos}
\email{pstavrin@math.uoa.gr}
\affiliation{Department of Mathematics, National and Kapodistrian University of 
Athens,	Panepistimiopolis 15784, Athens, Greece}

\author{A. Triantafyllopoulos}
\email{alktrian@phys.uoa.gr}
\affiliation{Section of Astrophysics, Astronomy and Mechanics, Department of 
Physics, National and Kapodistrian University of Athens, Panepistimiopolis 15784, Athens, Greece}


\begin{abstract}
In this article, we review some aspects of gravitational field and cosmology based on Finsler and Finsler-like generalized metric structures. The geometrical framework of these spaces allows further  investigation of locally-anisotropic phenomena  related to the gravitational field and  cosmological considerations, e.g  the extracted geodesics, deflection of light, Finsler-Einstein gravitational field equations , the Friedmann equations and the Raychaudhuri  equations include extra anisotropic  terms that in the Riemannian framework of General Relativity (GR) are not interpreted. This approach gives us the opportunity to extend the research with more degrees of freedom on the tangent bundle of  a spacetime manifold. In the above mentioned generalizations omitting the extra anisotropic terms we recover the framework of GR. In addition,  we study the gravitational Magnus effect in a generalized metric framework.

 Based on this approach, we describe further properties of Finsler-Randers (FR) and  Schwarzschild Finsler Randers (SFR)
 cosmological  models which are useful for the description and evolution of the universe.
\end{abstract}


\maketitle

\section{Locally-anisotropic structure of the gravitational field  and Finsler cosmology}

In the framework of General Relativity, the gravitational field is described by the Riemannian geometry. The theory of general relativity  describes with high accuracy  the observed gravitational effects between masses resulting from their warping of  spacetime and it  also predicts novel effects of gravity, such as gravitational waves, deflection of light, black holes and an effect of gravity on time known as gravitational time dilation \cite{Carroll,Wald:1984}. Many of these predictions have been confirmed by experiment or observation,  most recently the gravitational waves in a framework of an homogeneous and isotropic space-time. However, taking into account the abundance and nature of dark energy and dark matter, the nature of inflation, cosmological tensions such as the $H_0$ and $S_8$, the possible values of local anisotropy in the evolution of the universe, as well as the theoretical problems of the cosmological constant \cite{Bamba:2024,Abbot-et-al:2016,Abbot-et-al:2017,Farnes:2018,Cruz:2023lmn} the validity range of general relativity might be restricted.  There are plethora of cosmological models in the literature only some of them are viable for the description and the dynamical evolution  of the universe. Particular emphasis is placed on studying models that  are different in nature of geometry, in terms of introducing new fields and hence more degrees of freedom, and in terms of modifying the cosmological constant so that it can evolve over time \cite{Papagiannopoulos:2020mmm,Capozziello:2011et}. These models are studied in terms of their dynamical behavior through the critical point analysis, in which the different points refer to different cosmological eras, through the growth of matter perturbation analysis where one can compare the model’s clustering of matter to that that we observe today and finally where it is possible by reaching exact solutions for the models for specific potentials \cite{Khyllep:2021pcu,Basilakos:2019dof}.

Modified theories of gravity extend the form of general relativity through various methods, leading to different field equations and thus to different cosmological consequences. They play an essential role and contribute to modern cosmology, providing a foundation for the current understanding of physical phenomena of the Universe. Models of alternative gravity are extremely useful in order to provide information on whether these models can be consistent with the observed universe in terms of dynamical behavior, cosmic history as well as matter content. Furthermore, the combination of these tools can significantly constrain valid models or model scenarios that are consistent with a description of the universe.
In a more extended framework the gravitational field can be interpreted as the metric of spacetime, determining a force-field which contains the motion of  elementary particles of spacetime \cite{Kapsabelis:2022bue}.  GR predicts that curvature is produced not only by the distribution of mass-energy but also by its motion. This consideration reveals the Finslerian geometrical character of spacetime. In this framework, Relativity violations are arising from breaking the Lorentz symmetry caused by hypothetical backround fields that influence the metric, curvature, geodesics and the causal null cone \cite{Kostelecky:2011qz,Kouretsis:2008ha,Pfeifer:2011}.  This consideration is related to the local anisotropy which affects any gravitational {phenomena} and is incorporated in the total structure of the universe depending on the era of cosmological evolution. {Different types of cosmological models can describe the time evolution of the universe}. Modified gravitational theories with additional terms on the Friedmann equations {can result} to the appearance of dark matter and effective dark energy sectors. Finsler and Finsler-like geometries  are used to describe field equations, FRW and Raychaudhuri equations, geodesics and dark matter and dark energy effects. These types of geometries depend on position and velocity (direction/scalar) coordinates \cite{Stavrinos:2012ty,Hohmann:2020mgs,Vacaru:2012,Vacaru:2010fc,Stavrinos:2006rf,Triantafyllopoulos:2024xct}. In the framework of applications of Finsler geometry, many works in different directions of geometrical and physical structures have contributed to the extension of research for theoretical and observational approaches during the last years. We cite some recent references from the literature of the applications of Finsler geometry \cite{Kostelecky:2011qz,Caponio:2017lgy,Bubuianu:2018qsq,Pfeifer:2019wus,Javaloyes:2018lex,Hohmann:2020mgs,Caponio:2020ofw,Triantafyllopoulos:2020vkx,Konitopoulos:2021eav,Stavrinos:2021ygh,Hohmann:2021zbt,Javaloyes:2022fmp,Heefer:2022sgt,Bubuianu:2022nef,Savvopoulos:2023qfh,Hama:2022vob,Hama:2023bkl,Vacaru:2010fc,Li:2017vuc,Roopa:2020,Stavrinos:2021ygh,Angit:2022lfu,Nekouee:2022dtl,Zhu:2023kjx,Stavrinos:2021ygh}.
  
In the first period of development of applications of Finsler geometry to Physics especially to General Relativity remarkable works were given by G. Randers \cite{Randers1941} and J. Horvath \cite{Horvath1950}. Later, Einstein’s field equations are formulated by J. Horvath, Y. Takano \cite{Takano1968} and S. Ikeda \cite{Ikeda:1980} in the Finslerian framework. In these studies, the field equations had been considered without calculus of variations. G. S. Asanov researched the Finslerian gravitational field by using Riemannian osculating methods and derived Einstein field equations with variational principle \cite{Asanov1983}.
By introducing a vector field in the metric structure of spacetime the geometry can be changed. The gravitational field is attributed to a Sasaki-type metric on the tangent bundle of pseudo-Finsler or Finsler-like geometries which is imprinted in the context of locally-anisotropic energy-momentum tensor \cite{Kapsabelis:2023khh}. Because of dynamical coordinates, additional degrees of freedom are taken into account which can contribute in further understanding of the evolution and acceleration of the universe. In this approach Lorentz invariant violation (LV) can appear and the local-anisotropy can be represented by a vector field \cite{Kostelecky:2008be}. The production of LV can be diluted to thermal energy and a large amount of entropy. A basic kind of Finler space is the  Finsler-Randers (FR) metric space. This type of space and the induced cosmological asymmetric model is of special interest since the field equations include an extra geometrical-{dynamical} term that acts as an anisotropic dark energy-fluid {playing the role} of the  varying cosmological constant $\Lambda$ \cite{Papagiannopoulos:2020mmm}. This model can also describe an asymmetry between past and future which comes from matter collapses \cite{Oppenheimer:1939} under gravitation which is the source of very weak radiation which is called Hawking radiation. The FR cosmological model contains in each point two metric structures, the Riemannian and the Finslerian one so it can be considered as a direction-depended motion of the Riemannian FRW-model .
Studying the dynamics in varying vaccuum Finsler-Randers cosmology {were we have considered also interactions} we find new eras in the cosmological history provided by the geometrodynamical terms \cite{Papagiannopoulos:2020mmm}.  Schwarzschild Finsler Randers (SFR) spacetime shows a motion of the Schwarzschild model with a produced work which comes from the second (one-form) \cite{Triantafyllopoulos:2020vkx,Triantafyllopoulos:2020ogl}. This consideration gives interactions between a force field $F_a$ and the vector field $y_a$ which is incorporated in the metric structure of Finsler space. This type of metric provides like as in the case of FR model, a dynamical effective structure in the spacetime and simultaneously produces a bimetric gravity field theory that encompasses  SFR-model generalizing the classic Schwarzschild spacetime by introducing a timelike covector in the metric structure which is specified by the solution of the generalized Einstein equations of the SFR model. It produces local anisotropy and may introduce Lorentz violating effects. In addition, we provide the S-anisotropic curvature which plays a significant role extending the framework of Kretschmann curvature invariant $K_V$ giving the dependence with S-curvature.
By this relation we get more information for the singularities than of $K_H$ horizontal Kretschmann curvature invariant in which $K_H^{(GR)}$ = $K_H^{(SFR)}$. We calculate the gravitational redshift and the photonsphere in our case \cite{Triantafyllopoulos:2020vkx,Triantafyllopoulos:2020ogl}.
We prove that the gravitational redshift predicted by our model remains invariant compared {to} the one of GR. In the case of photonsphere, we find infinitesimal deviations from GR which may be {attributed} to the small anisotropic perturbations coming from Lorentz violation effects giving an extension to the results of general relativity including locally anisotropic extensions to the observable phenomena \cite{Kapsabelis:2022bue}.

\section{Varying  metrics in the universe}

There is not a unique metric investigating the structure of the evolution of the universe since different cosmological regions depend on the distribution of matter and energy as well as the fluctuation of  CMB.  In addition, taking into account a primordial magnetic field on the metric structure of the tangent bundle spacetime, the description requires an anisotropic character including additional curvatures in the geometric stucture. For instance, near a black hole strong magnetic fields play an important role, a case is the super massive black hole  M 87* \cite{EHT:2023ujh}.
   The description of the universe based on the metrics which can be changed in different eras during its expansion, e.g, in empty space,
the de-Sitter model or Minkowski metric  for the description of a part of spacetime without the presence of matter. Riemannian metric 
is closely related to the standard General Relativity and the isotropic evolution. On the other hand, locally-anisotropic descriptions and time-asymmetry of cosmological phenomena can be studied in the framework of Finsler and Finsler-like geometries of spacetime which constitute a natural metric generalization of the Riemannian geometry. In this framework, a cosmological model in the FR space has been introduced in \cite{stavrinos2005,Stavrinos:2006rf}.

In FR space, the energy-momentum tensor describes the structure of spacetime and it measures  the curvature in an anisotropic space where the horizontal and vertical energy momentum tensors describe how the energy and momentum are distributed in the direction/velocity of spacetime \cite{stavrinos2005,Stavrinos:2006rf,Triantafyllopoulos:2018bli}.

\section{Generalized Einstein-Finsler-like field equations}

Different investigations  of generalized Einstein field equations were derived for the aforementioned spaces in the framework of a tangent bundle.  Additionally, Lorentz invariance violation in Finsler/Finslerlike spacetime and in Finsler cosmology in very special relativity has been studied in the works \cite{Kostelecky:2008be,Kostelecky:2011qz,Gibbons:2007iu}.

Generalized Einstein field equations have been explored  on the Lorentz tangent bundle of the Finsler,  and Finsler-like spacetimes as well as  for an osculating gravitational approach in the Finsler cosmology in which the second variable $y(x)$ is a tangent vector/ field \cite{Kapsabelis:2023khh,Stavrinos:2006rf}. These equations  constitute the base for deriving of generalized Friedmann equations which can include dark matter or dark energy terms \cite{Kouretsis:2008ha,Konitopoulos:2021eav}.
Investigations of the extended Friedmann equations in Finsler spaces with extra internal degrees of freedom and dynamical analysis (critical points) provide a better understanding of the dynamical properties of the Finsler–Randers cosmological models. To this end, articles in the framework of the weak field and pp-waves in Finsler spacetime can be found in \cite{Kostelecky:2011qz,AlanKostelecky:2012yjr,Triantafyllopoulos:2020ogl,Triantafyllopoulos:2018bli,Fuster:2015tua}
and potentially they can be used in order to test the performance of the Finslerian gravitational theory against current observations of gravitational waves.

In the generalized framework of a Lorentz tangent bundle, the field equations for the metric can be derived from a Hilbert-like action \cite{Triantafyllopoulos:2020ogl}:
    \begin{equation}\label{Hilbert like action}
        K = \int_{\mathcal N} d^8\mathcal U \sqrt{|\Gd|}\, \R + 2\kappa \int_{\mathcal N} d^8\mathcal U \sqrt{|\Gd|}\,\mathcal L_M
    \end{equation}
    for some closed subspace $\mathcal N\subset TM$, where $|\Gd|$ is the absolute value of the metric determinant, $\mathcal L_M$ is the Lagrangian of the matter fields, $\kappa$ is a constant and
    \begin{equation}
        d^8\mathcal U = \de x^0 \wedge \ldots \wedge\de x^3 \wedge \de y^0 \wedge \ldots \wedge \de y^3
    \end{equation}
{ where the 8-parallelepiped  $d^{8}\mathcal{U}$ is considered an oriented compact element of volume.}

The solution for a stationary action is given by the field equations:
\begin{gather}
	{\overline R_{\mu\nu}} - \frac{1}{2}({\overline R}+{S})\,{g_{\mu\nu}} + \left(\delta^{(\lambda}_\nu\delta^{\kappa)}_\mu - g^{\kappa\lambda}g_{\mu\nu} \right)\left(\mathcal D_\kappa\mathcal T^\beta_{\lambda\beta} - \mathcal T^\gamma_{\kappa\gamma}\mathcal T^\beta_{\lambda\beta}\right) = \kappa {T_{\mu\nu}} \label{feq1}\lin
	{S_{\alpha\beta}} - \frac{1}{2}({\overline R}+{S})\,{v_{\alpha\beta}}  + \left(v^{\gamma\delta}v_{\alpha\beta} - \delta^{(\gamma}_\alpha\delta^{\delta)}_\beta \right)\left(\mathcal D_\gamma C^\mu_{\mu\delta} - C^\nu_{\nu\gamma}C^\mu_{\mu\delta} \right) = \kappa {Y_{\alpha\beta}} \label{feq2}\lin
	g^{\mu[\kappa}\pdot{\alpha}L^{\nu]}_{\mu\nu} +  2 \mathcal T^\beta_{\mu\beta}g^{\mu[\kappa}C^{\lambda]}_{\lambda\alpha} = \frac{\kappa}{2}\mathcal Z^\kappa_\alpha \label{feq3}
\end{gather}
with
\begin{align}
	T_{\mu\nu} \equiv -\frac{2}{\sqrt{|\Gd|}}\frac{\Delta\left(\sqrt{|\Gd|}\,\mathcal{L}_M\right)}{\Delta g^{\mu\nu}}\label{em1}\\
	Y_{\alpha\beta} \equiv -\frac{2}{\sqrt{|\Gd|}}\frac{\Delta\left(\sqrt{|\Gd|}\,\mathcal{L}_M\right)}{\Delta v^{\alpha\beta}} \label{em2}\\
	\mathcal Z^\kappa_\alpha \equiv -\frac{2}{\sqrt{|\Gd|}}\frac{\Delta\left(\sqrt{|\Gd|}\,\mathcal{L}_M\right)}{\Delta N^\alpha_\kappa}\label{em3}
\end{align}
and $\kappa$ constant.

The concepts of generalized Ricci curvature $\overline R_{\mu\nu}, S_{\alpha\beta}, \overline R, S, \mathcal R$, torsion $T^\alpha_{\lambda\beta}, C^\mu_{\nu\gamma}$, covariant derivative $\mathcal D_\alpha$, energy-momentum $T_{\mu\nu}, Y_{\alpha\beta}, \mathcal Z^\kappa_\alpha$ and metric $g_{\mu\nu}, v_{\alpha\beta}, \mathcal G$ contained in relations (\ref{Hilbert like action}-\ref{em3}) are elaborated in \cite{Triantafyllopoulos:2018bli,Triantafyllopoulos:2020ogl}.

\section{The cosmological models FR and SFR}

We will give some additional remarks  for the models of FR and SFR  that can reveal  further properties.
As we mentioned before, the introduction of a vector field in the metric structure  of spacetime anisotropically affects all the geometrical and physical concepts as the geodesics, curvature, gravitational filed equations, energy momentum tensor, Friedmann equations  e.t.c,  including  anisotropic terms which further extend the investigation with more degrees of freedom in the evolution and acceleration of the universe.
In addition, in the dynamical analysis of FR cosmological model {we have }found solutions which accommodate cosmic acceleration and under specific conditions they provide de-Sitter points as stable late-time attractors \cite{Papagiannopoulos:2020mmm,Papagiannopoulos:2017whb}.
The FR  cosmological model is based on the Finslerian geometry on the total space of the tangent bundle of spacetime, it is  a natural generalization of the standard Riemannian space. In this framework, the density $\rho$ and the pressure  $p$ of a cosmological perfect fluid are reduced  on the tangent bundle of spacetime because of increasing of the volume in more dimensions  \cite{Basilakos:2013hua}. The relation between densities and pressures of the fluid on the background spacetime and its Finslerian tangent bundle $\rho^{(f)}$, $p^{(f)}$  are given in the following forms,
\begin{align}
    \rho^{(f)} & = \frac{\alpha^2}{F^2}\rho \\
    p^{(f)} & = \frac{ \alpha}{F}p 
\end{align}
where  $ \alpha $ , $ F $ represent the arc lengths of Riemann and FR spaces respectively, with $ |\alpha| < |F| $.

\subsection{Aspects on the FR model}
The Finsler-Randers (FR) spacetime was first proposed by Randers \cite{Randers1941} and constitutes a significant class of Finslerian spacetime. The metric function in this space for a spinning particle takes the form
\begin{equation}\label{lagrangian}
    F(x,y) = (-a_{\mu\nu}(x)y^{\mu}y^{\nu})^{1/2} + F_{\alpha}y^{\alpha}    
\end{equation}
where $F_{\alpha} = \Phi_\alpha + M_\alpha$ is considered to be small, i.e. $||F_{\alpha}||\ll 1$, $\Phi_\alpha$ is a weak anisptropic field and $M_\alpha$ is a weak Magnus force field, $y^{\alpha}=\frac{dx^{\alpha}}{d\tau}$ and $a_{\mu\nu}(x)$ is a Riemannian metric with a Lorentzian signature $(-,+,+,+)$  and the indices $\mu, \nu, \alpha$ take the values $0,1,2,3$. The geodesics of this space are produced by \eqref{lagrangian} by means of the Euler-Lagrange equations. If we assume that $F_{\alpha}$ represents a force field $f_{\alpha}$ and $y^{\alpha}$ is the 4-velocity $d x^{\alpha}$ then $f_{\alpha}dx^{\alpha}$ can represent the spacetime analog of infinitesimal work produced by the anisotropic force field $f_{\alpha}$, therefore we write equation \eqref{lagrangian} as
\begin{equation}\label{lagrangian2}
    F(x,dx) = \left(-a_{\mu\nu}(x)dx^{\mu}dx^{\nu}\right)^{1/2} + f_{\alpha}dx^{\alpha}    
\end{equation}
The integral $\int_{a}^{b}F(x,dx)$ represents the total spacetime work of the force field $f_{\alpha}$ along a particle's path.

The length of a curve $c$ in the FR space is given by 
\begin{equation}
    l(c) = \int^{1}_{0}F(x,\dot{x})d\tau    
\end{equation}
where $\dot{x} = \frac{dx}{d\tau}$ and $\tau$ is an affine parameter.

An FR cosmological model has been introduced and studied in the literature, in \cite{stavrinos2005,Stavrinos:2006rf}. In this model, the Riemannian metric $a_{\mu\nu}(x)$ in \eqref{lagrangian2} is substituted with the classic FRW metric: 
\begin{equation}
    a_{\mu\nu}(x) = \mathrm{diag}\left[-1,\frac{a^2}{1-\kappa r^{2}},a^{2}r^{2}, a^{2}r^{2}\sin^{2}\theta\right]   
\end{equation}
therefore, rel. \eqref{lagrangian2} is an extension of the classic FRW cosmology with the anisotropy-inducing term $f_{\alpha}dx^{\alpha}$. The resulting model is a Finsler-Randers cosmology. An interpretation of rel. \eqref{lagrangian2} can be that the FR spacetime shows a motion of the FRW model with a produced work which comes from the second term (one-form).
 
Anisotropy in the distribution of matter means a gravitational potential and can be connected to the dark energy.  The interaction of the gravitational field with  matter can be characterized by potential energy which can be ‘work’, that means it can be converted to kinetic energy. In addition, the  gravitational field in a FR space includes by virtue of the second term, angular momentum which  is imprinted on the geodesics of a FR space. This consideration leads  to rotational geodesics that are useful  to describe of cosmological phenomena related to black holes or accretion disks in which material of gas or dust affect the original geodesics of an astronomical object \cite{Abramowicz:2011xu}. The FR cosmological model is a viable model arising from the results of dynamical analysis of the model \cite{Papagiannopoulos:2020mmm,Papagiannopoulos:2017whb}.
Investigating the gravitational field on a FR spacetime we notice that it carries energy and momentum in the form of gravitational waves 
that it would also be kinetic energy since it encompasses the motion. The energy comes from the metric of FR, $dt/d\lambda=E=y_0 =P_0$.
  For a photon, we get $ v_0  =dt/d\lambda=\omega_0  /a$  ,  where $y_0$ denotes the time component of the second term of the metric and $P ,v, a, \lambda, \omega$  denote the momentum, velocity, scale factor, an affine parameter and the frequency respectively.
The geodesic equation in the space after the above mentioned remarks takes the form

\begin{equation}
    \frac{dP^\kappa}{d\lambda}  + L^\kappa_{\rho\sigma} P^\rho P^\sigma = 0
\end{equation}
where $P^\kappa = dx^\kappa/d\lambda $  and  $ L^\kappa_{\rho\sigma} $ represent the connection coefficients of the FR space.

From the Euler-Lagrange equations
\begin{equation}
	\frac{d}{d\tau}\frac{\partial L}{\partial \dot{x}^{\mu}}=\frac{\partial L}{\partial x^{\mu}}    
\end{equation}
we find the equations for the geodesics:
\begin{equation}
	\ddot{x}^{\lambda}+\Gamma^{\lambda}_{\mu\nu}\dot{x}^{\mu}\dot{x}^{\nu}+g^{\kappa\lambda}\Phi_{\kappa\mu}\dot{x}^{\mu}=0  
\end{equation}
where $\Gamma^{\lambda}_{\mu\nu}$ are the Christoffel symbols of Riemann geometry, $\dot{x}^{\mu}=\frac{dx^{\mu}}{d\tau}$ and $\Phi_{\kappa\mu}=\partial_{\kappa}A_{\mu}-\partial_{\mu}A_{\kappa}$ and $A_{\mu}$ is the solution rel.\eqref{Asolution}. We notice that from the definition of $\Phi_{\kappa\mu}$ we get a rotation form of geodesics. If $A_{\mu}$ is a gradient of a scalar field, $A_{\mu}=\pder{\Phi}{x^{\mu}}$ then $\Phi_{\kappa\mu}=0$ and the geodesics of our model are identified with the Riemannian ones.

\subsection{Elements of the SFR model}
A natural framework for the study of a SFR cosmological model  is the Lorentz tangent bundle of a spacetime manifold with Schwarzschild metric \cite{Triantafyllopoulos:2020vkx}. Solving the generalized field equations for the perturbed metric, we get the dynamics of the covector $A_\gamma$. The derived solution is called the Schwarzschild–Finsler-Randers spacetime. 
The metric of this extended spacetime takes the form
\begin{equation}
    \mathcal{G} = g_{\mu\nu}(x,y)\,\mathrm{d}x^\mu \otimes \mathrm{d}x^\nu + v_{\alpha\beta}(x,y)\,\delta y^\alpha \otimes \delta y^\beta \label{bundle metric}
\end{equation}
The horizontal part $g_{\mu\nu}$ of the metric \eqref{bundle metric} is the classic Schwarzschild metric:
\begin{equation}\label{Schwarzchild}
    g_{\mu\nu}\de x^\mu  \de x^\nu = -\left(1-\frac{R_s}{r}\right)dt^2 + \frac{dr^2}{1-\frac{R_s}{r}} + r^2 d\theta^2 + r^2 sin^{2}\theta d\phi^2
\end{equation}
where $R_s=2GM$ is the Schwarzschild radius ($c=1$). 
	
We assume a Finsler function $ F $ of Randers type:
\begin{equation}\label{RandersL}
    F(x,y) = \sqrt{-g_{\alpha\beta}(x)y^\alpha y^\beta} + A_\gamma(x) y^\gamma
\end{equation}
where $g_{\alpha\beta}$ is the Schwarzschild metric and $A_{\gamma}(x)$ is a covector which is determined by the field equations of the generalized metric. 
The metric tensor $v_{\alpha\beta}$ is given by:
\begin{equation}\label{vRandersmetric}
    v_{\alpha\beta}= -\frac{1}{2}\frac{\partial^{2}F^2}{\partial y^{\alpha}\partial y^{\beta}}  
\end{equation}
Solving the field equations \eqref{feq1}-\eqref{feq3} gives the one-form $A_\gamma$:
\begin{equation}\label{Asolution}
    A_\gamma(x) = \left[\tilde A_0 \left|1-\frac{R_S}{r} \right|^{1/2}, 0, 0, 0 \right]
\end{equation}
with $\tilde A_0$ a constant. This is a timelike covector since to second order in $A_\gamma$ we get $g^{\alpha\beta}A_\alpha A_\beta = -(\tilde A_0)^2 < 0$.
\\\\
{\it Remark:} Substituting the metric $v_{\alpha\beta}$ with an angular metric $\phi_{\alpha\beta}$ \cite{rund59}, we can study spinning phenomena in a space of the form \eqref{bundle metric}.

\section{Gravitational Magnus effect in FR-space}

The Magnus effect is the force exerted perpendicular to the motion of a spinning object and its rotation axis  moving in a fluid. In general relativity, an analogous effect exists for a spinning compact object moving through the cosmological fluid  as a result of gravitational interactions \cite{Tsuji:1985,Costa:2018,Wang:2024cej}. 
The standard Magnus force $\mathbf M$ has a direction in $v\times\omega$, where $v$ is the velocity of the fluid relative to the body and $\omega$ the spin angular velocity of the body. Taking into account  the fluid density  $\rho$ the Magnus force  can be defined  by  $\mathbf M =  k\rho( v\times\omega )$, with $k$ a factor that depends on the type of fluid.
 This gravitational phenomenon can also be  connected with  a spinning Kerr black hole moving at relativistic velocities \cite{Okawa:2014sxa}.
 \begin{figure}[h]
    \centering
    \includegraphics[width=400px]{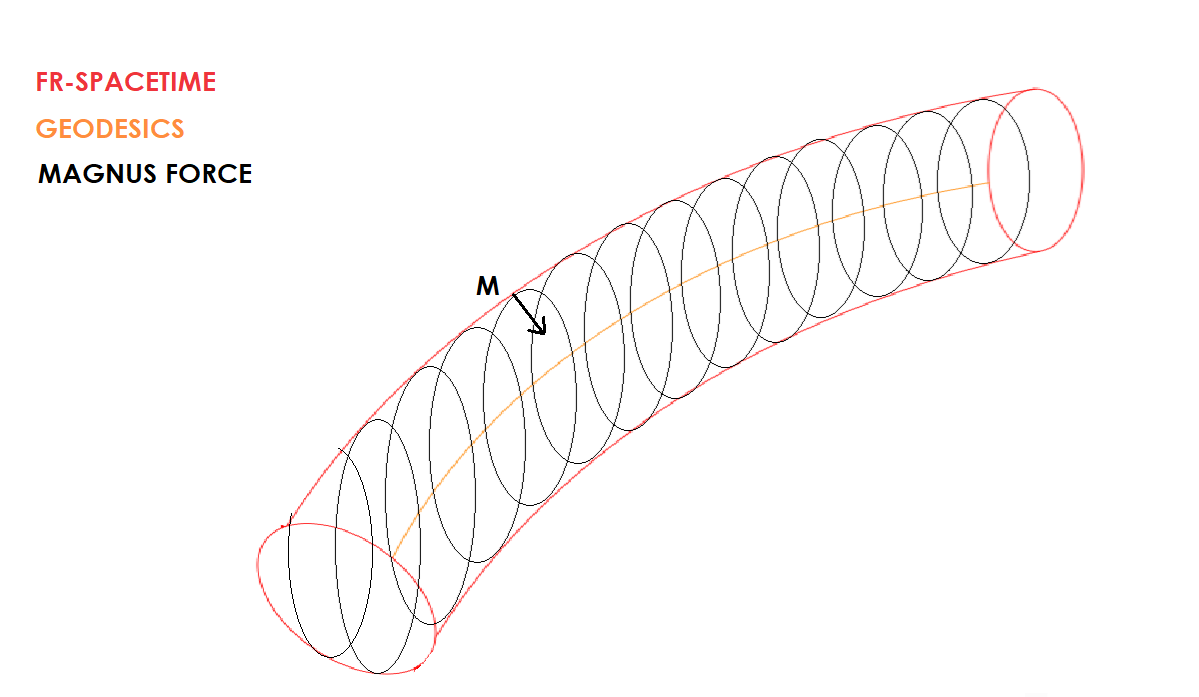}
    \caption{Spinning movement of a particle under the effect of Magnus force in FR space}
    \label{fig:magnus_force}
\end{figure}

The Magnus phenomenon can be generalized  in a FR spacetime in which the extracted geodesics are rotated, Figure \ref{fig:magnus_force} , therefore, a spinning particle or body  moving in a cosmological fluid or viscuus and dark matter halo accepts a  vertical force which is  the Magnus force and it and can be intrinsically incorporated  in the metric structure of the gravitational field as it is mentioned in rel. \eqref{lagrangian}.
The geodesics on FR space when are influenced from the Magnus field,  are given in the form
\begin{equation}\label{msfr-geodesics}
    \ddot{x}^\kappa + \Gamma^\kappa_{\mu\nu} \dot x^\mu \dot x^\nu + g^{\kappa\mu} \left(\Phi_{\lambda\mu} + M_{\lambda\mu} \right) x^\lambda = 0
\end{equation} 
where  $\Gamma^\kappa_{\mu\nu}$  are the Christoffel symbols of Riemann geometry,
$\dot x^\mu  = dx^\mu/d\tau$, $\tau$ is the proper time, $\Phi_{\kappa\mu} = \partial_\kappa A_\mu - \partial_\mu A_\kappa$, $\Phi_{\kappa\mu}$ represents the stress tensor of anisotropy caused by the potential vector field $A_\mu$ of anisotropy.   In a similar way, the stress tensor $ M_{\kappa\mu} $ of  Magnus force $\mathbf M$ can be defined. 
We can notice  from the curl $ \Phi_{\kappa\mu} $ that we get a rotation form of geodesics. If $A_\mu$ is a gradient of a scalar field, $A_\mu = \partial \phi / \partial x^\mu$ then $\Phi_{\kappa\mu} = 0$ and the geodesics of our model are identified with the Riemannian one. The term  $M_{\kappa\lambda}$ denotes  the Magnus force field which is caused when the path of the particle is rotated during its motion, this phenomenon can be derived by extra dust or clouds of fluids that affect the geodesic motion of a particle or of a body .  In the case we consider the dark energy with $\rho=-p$, with $\rho$ and $p$ the density and the pressure of the cosmological fluid , the Magnus effect is zero, $M_{\kappa\lambda}  = 0$ \cite{Costa:2018} and the form of geodesics  is reduced in the standard one of  FR space \cite{Stavrinos:2016xyg}. It is obvious that when the anisotropic terms $\Phi$ and $\mathbf M$ are zero, Eq. \eqref{msfr-geodesics} represents Riemannian geodesics.
We can see  that the geodesic equation \eqref{msfr-geodesics}  encompasses  more amount of anisotropy because of  the Magnus force $\mathbf M $ which exerts an additional vertical force on a spinning particle / body moving in the gravitational field of the FR space giving further deviation in their form.

In the original form of a FR space, the covector $A_\mu$  represents an electromagnetic potential \cite{Randers1941} and the space encompasses gravito-electromagnetic curvature which consists of two parts, one of them is  Riemannian and the second one contains electromagnetic terms \cite{Stavrinos:2012kv,asanov1985}.  In that case, the FR space can give a generalized physical background describing and investigating  gravitational and astrophysical phenomena connected with  equations of motion for spinning bodies and Magnus effect. This will be an object of further study in the near future.

\section{Main Points}

Based on the above mentioned considerations, we conclude that it is possible to extend
gravitational and cosmological models with isotropic and locally-anisotropic phenomena.  In addition, we study the gravitational Magnus effect in a generalized metric
framework of an FR space. The geometrical extension of the cosmological model of GR with the corresponding FR and SFR models of Modified Gravity, opens up windows in this direction of research for testing gravitational theories with generalized metric structure through further experiments.


\begin{thebibliography}{99}

\bibitem{Carroll}
S.~Carroll,
{\it Spacetime and Geometry, An Introduction to General Relativity},
Pearson Education Inc, Addison Wesley, (2004)

\bibitem{Wald:1984}
R.~Wald,
{\it General Relativity},
Chicago University Press (1984)


\bibitem{Bamba:2024}
K.~Bamba,
Universe {\bf 10} (2024), 144

\bibitem{Abbot-et-al:2016}
B.~P.~Abbott et al. [LIGO Scientific Collaboration and Virgo Collaboration]
Phys. Rev. Lett. {\bf 116} (2016), 061102 

\bibitem{Abbot-et-al:2017}
B.~P.~Abbott et al. [LIGO Scientific Collaboration and Virgo Collaboration]
Phys. Rev. Lett. {\bf 119} (2017), 161101 

\bibitem{Farnes:2018}
J.~S.~Farnes
Astronomy and Astrophysics {\bf 620} (2018)

\bibitem{Cruz:2023lmn}
J.~S.~Cruz, F.~Niedermann and M.~S.~Sloth,
JCAP \textbf{11} (2023), 033


\bibitem{Papagiannopoulos:2020mmm}
G.~Papagiannopoulos, S.~Basilakos, A.~Paliathanasis, S.~Pan and P.~Stavrinos,
Eur. Phys. J. C \textbf{80} (2020) no.9, 816


\bibitem{Capozziello:2011et}
S.~Capozziello and M.~De Laurentis,
Phys. Rept. \textbf{509} (2011), 167-321


\bibitem{Khyllep:2021pcu}
W.~Khyllep, A.~Paliathanasis and J.~Dutta,
Phys. Rev. D \textbf{103} (2021) no.10, 103521

\bibitem{Basilakos:2019dof}
S.~Basilakos, G.~Leon, G.~Papagiannopoulos and E.~N.~Saridakis,
Phys. Rev. D \textbf{100} (2019) no.4, 043524


\bibitem{Kapsabelis:2022bue}
E.~Kapsabelis, P.~G.~Kevrekidis, P.~C.~Stavrinos and A.~Triantafyllopoulos,
Eur. Phys. J. C \textbf{82} (2022) no.12, 1098


\bibitem{Kostelecky:2011qz}
A.~Kostelecky,
Phys. Lett. B \textbf{701} (2011), 137-143

\bibitem{Kouretsis:2008ha}
A.~P.~Kouretsis, M.~Stathakopoulos and P.~C.~Stavrinos,
Phys. Rev. D \textbf{79} (2009), 104011

\bibitem{Pfeifer:2011}
C.~Pfeifer and M.~N.~R.~Wohlfarth
Phys. Rev. D {\bf 84}, 044039 

\bibitem{Stavrinos:2012ty}
P.~C.~Stavrinos and S.~I.~Vacaru,
Class. Quant. Grav. \textbf{30} (2013), 055012

\bibitem{Hohmann:2020mgs}
M.~Hohmann, C.~Pfeifer and N.~Voicu,
Universe \textbf{6} (2020) no.5, 65

\bibitem{Vacaru:2012}
S.~Vacaru,
Int. J. Mod. Phys. D 21 (2012) 1250072

\bibitem{Vacaru:2010fc}
S.~I.~Vacaru,
Phys. Lett. B \textbf{690} (2010), 224-228

\bibitem{Stavrinos:2006rf}
P.~C.~Stavrinos, A.~P.~Kouretsis and M.~Stathakopoulos,
Gen. Rel. Grav. \textbf{40} (2008), 1403-1425

\bibitem{Stavrinos:2012kv}
P.~Stavrinos,
Gen. Rel. Grav. \textbf{44} (2012), 3029-3045

\bibitem{Triantafyllopoulos:2024xct}
A.~Triantafyllopoulos, E.~Kapsabelis and P.~C.~Stavrinos,
Universe \textbf{10} (2024) no.1, 26

\bibitem{Caponio:2017lgy}
E.~Caponio and G.~Stancarone,
Class. Quant. Grav. \textbf{35} (2018) no.8, 085007

\bibitem{Bubuianu:2018qsq}
L.~Bubuianu and S.~I.~Vacaru,
Annals Phys. \textbf{404} (2019), 10-38

\bibitem{Pfeifer:2019wus}
C.~Pfeifer,
Int. J. Geom. Meth. Mod. Phys. \textbf{16} (2019) no.supp02, 1941004

\bibitem{Javaloyes:2018lex}
M.~\'A.~Javaloyes and M.~S\'anchez,
RACSAM \textbf{114} (2020), 30


\bibitem{Caponio:2020ofw}
E.~Caponio and A.~Masiello,
Universe \textbf{6} (2020) no.4, 59

\bibitem{Triantafyllopoulos:2020vkx}
A.~Triantafyllopoulos, S.~Basilakos, E.~Kapsabelis and P.~C.~Stavrinos,
Eur. Phys. J. C \textbf{80} (2020) no.12, 1200

\bibitem{Konitopoulos:2021eav}
S.~Konitopoulos, E.~N.~Saridakis, P.~C.~Stavrinos and A.~Triantafyllopoulos,
Phys. Rev. D \textbf{104} (2021) no.6, 064018

\bibitem{Li:2017vuc}
X.~Li and H.~N.~Lin,
Eur. Phys. J. C \textbf{77} (2017) no.5, 316

\bibitem{Roopa:2020}
M.~K.~Roopa and S.~K.~Narasimhamurthy,
Palestine Journal of Mathematics \textbf{9} (2020) no.2, 957–968 

\bibitem{Stavrinos:2021ygh}
P.~Stavrinos and S.~I.~Vacaru,
Universe \textbf{7} (2021) no.4, 89

\bibitem{Angit:2022lfu}
S.~Angit, R.~Raushan and R.~Chaubey,
Pramana \textbf{96} (2022) no.3, 123

\bibitem{Nekouee:2022dtl}
Z.~Nekouee, S.~K.~Narasimhamurthy, H.~M.~Manjunatha and S.~K.~Srivastava,
Eur. Phys. J. Plus \textbf{137} (2022) no.12, 1388

\bibitem{Zhu:2023kjx}
J.~Zhu and B.~Q.~Ma,
Symmetry \textbf{15} (2023) no.5, 978


\bibitem{Hohmann:2021zbt}
M.~Hohmann, C.~Pfeifer and N.~Voicu,
J. Math. Phys. \textbf{63} (2022) no.3, 032503

\bibitem{Javaloyes:2022fmp}
M.~\'A.~Javaloyes, M.~S\'anchez and F.~F.~Villase\~nor,
Universe \textbf{8} (2022) no.2, 93

\bibitem{Heefer:2022sgt}
S.~Heefer, C.~Pfeifer, J.~van Voorthuizen and A.~Fuster,
J. Math. Phys. \textbf{64} (2023) no.2, 022502

\bibitem{Bubuianu:2022nef}
L.~Bubuianu, D.~Singleton and S.~I.~Vacaru,
JHEP \textbf{05} (2023), 057

\bibitem{Hama:2022vob}
R.~Hama, T.~Harko and S.~V.~Sabau,
Eur. Phys. J. C \textbf{82} (2022) no.4, 385

\bibitem{Savvopoulos:2023qfh}
C.~Savvopoulos and P.~C.~Stavrinos,
Phys. Rev. D \textbf{108} (2023) no.4, 044048

\bibitem{Hama:2023bkl}
R.~Hama, T.~Harko and S.~V.~Sabau,
Eur. Phys. J. C \textbf{83} (2023) no.11, 1030

\bibitem{Randers1941}
G.~Randers,
Phys. Rev. {\bf 59} (1941) no.2, 195--199

\bibitem{Horvath1950}
J.~I.~Horváth,
Phys. Rev. {\bf 80} (1950) 901

\bibitem{Takano1968}
Y.~Takano,
Prog. Theor. Phys. {\bf 40} (1968) no.5, 1159-1180

\bibitem{Ikeda:1980}
S.~Ikeda,
Rep. Math. Phys. {\bf 18} (1980) 103

\bibitem{Asanov1983}
G.~S.~Asanov,
Foundations of Physics {\bf 13} (1983) no.5, 501-527

\bibitem{Kapsabelis:2023khh}
E.~Kapsabelis, E.~N.~Saridakis and P.~C.~Stavrinos,
[arXiv:2312.15552 [gr-qc]].
To be published in EPJC

\bibitem{Kostelecky:2008be} 
V.~A.~Kosteleck\'y and M.~Mewes,
{\it Astrophysical Tests of Lorentz and CPT Violation with Photons,}
Astrophys.\ J.\  {\bf 689} (2008) L1

\bibitem{Oppenheimer:1939}
J.~R.~Oppenheimer and H.~Snyder,
Phys. Rev. {\bf 56} (1939), 455

\bibitem{Triantafyllopoulos:2020ogl}
A.~Triantafyllopoulos, E.~Kapsabelis and P.~Stavrinos,
Eur. Phys. J. Plus \textbf{135} (2020) no.7, 557

\bibitem{EHT:2023ujh}
The Event Horizon Telescope Collaboration  \textit{et al.} [EHT],
Astrophys. J. Lett. \textbf{957} (2023) no.2, L20

\bibitem{stavrinos2005}
P.~C.~Stavrinos, 
Int. J. Theor. Phys. {\bf 44} (2005) 245-254

\bibitem{Triantafyllopoulos:2018bli}
A.~Triantafyllopoulos and P.~C.~Stavrinos,
Class. Quant. Grav. \textbf{35} (2018) no.8, 085011

\bibitem{Gibbons:2007iu}
G.~W.~Gibbons, J.~Gomis and C.~N.~Pope,
Phys. Rev. D \textbf{76} (2007), 081701

\bibitem{AlanKostelecky:2012yjr}
V.~Alan Kosteleck\'y, N.~Russell and R.~Tso,
Phys. Lett. B \textbf{716} (2012), 470-474

\bibitem{Fuster:2015tua}
A.~Fuster and C.~Pabst,
Phys. Rev. D \textbf{94} (2016) no.10, 104072

\bibitem{Basilakos:2013hua}
S.~Basilakos, A.~P.~Kouretsis, E.~N.~Saridakis and P.~Stavrinos,
Phys. Rev. D \textbf{88} (2013), 123510

\bibitem{Abramowicz:2011xu}
M.~A.~Abramowicz and P.~C.~Fragile,
Living Rev. Rel. \textbf{16} (2013), 1

\bibitem{Papagiannopoulos:2017whb}
G.~Papagiannopoulos, S.~Basilakos, A.~Paliathanasis, S.~Savvidou and P.~C.~Stavrinos,
Class. Quant. Grav. \textbf{34} (2017) no.22, 225008

\bibitem{rund59}
H.~Rund,
{\it The Differential Geometry of Finsler Spaces,}
Springer (1959), Berlin

\bibitem{Tsuji:1985}
Y.~Tsuji, Y.~Morikawa and O.~Mizuno,
ASME. J. Fluids Eng. \textbf{107} (1985) no.4, 484–488

\bibitem{Costa:2018}
L.~F.~O.~Costa, R.~Franco and V.~Cardoso,
Phys. Rev. D \textbf{98} (2018), 024026

\bibitem{Wang:2024cej}
Z.~Wang, T.~Helfer, D.~Traykova, K.~Clough and E.~Berti,
[arXiv:2402.07977 [gr-qc]].


\bibitem{Okawa:2014sxa}
H.~Okawa and V.~Cardoso,
Phys. Rev. D \textbf{90} (2014) no.10, 104040

\bibitem{Stavrinos:2016xyg}
P.~C.~Stavrinos and M.~Alexiou,
Int. J. Geom. Meth. Mod. Phys. \textbf{15} (2017) no.03, 1850039

\bibitem{asanov1985}
G. S. Asanov,
{\it Finsler Geometry, Relativity and Gauge theories,}
D. Reidel Publishing Company (1985), Dordrecht, Holland


\end{thebibliography}
\end{document}